\documentclass[10pt]{article}
\usepackage{epsfig}
\renewcommand{\author}[3]{\begin{center}
                           {\sc #1}\\
                           {#2}
                          \end{center}}
\renewcommand{\title}[1]{\begin{center}
                      {\Large {\bf #1}}
                      \end{center}}

 \def\la{\hbox{\raise.5ex\hbox{$<$}
     \kern-1.1em\lower.5ex\hbox{$\sim$}}}
 \def\ga{\hbox{\raise.5ex\hbox{$>$}
     \kern-1.1em\lower.5ex\hbox{$\sim$}}}

\begin{document}

\title{
Riemann Solvers in  General Relativistic Hydrodynamics
\vspace*{5mm}}

\author{J.M$^{\underline{\mbox{a}}}$ Ib\'a\~nez$^1$,
	M.A. Aloy$^1$, 
	J.A. Font$^2$, 
        J.M$^{\underline{\mbox{a}}}$ Mart\'{\i}$^1$,
	J.A. Miralles$^1$ and
	J.A. Pons$^1$
}
{
$^1$ Departamento de Astronom\'{\i}a y Astrof\'{\i}sica,
                  Universidad de Valencia, 46100 Burjassot (Valencia), Spain \\
$^2$ Max--Planck--Institut f\"ur Astrophysik,
                  Karl--Schwarzschild--Str.\,1, D--85740 Garching, Germany
}

\vspace{5mm}

\begin{abstract}

Our contribution concerns with the numerical solution of the 3D
general relativistic hydrodynamical system of equations within the framework
of the $\{3+1\}$ formalism. We summarize the theoretical ingredients which
are necessary in order to build up a numerical scheme based on the solution
of local Riemann problems. Hence, the full spectral decomposition of the
Jacobian matrices of the system, i.e., the eigenvalues and the right and left
eigenvectors, is explicitly shown.
An alternative approach consists in using any of the special relativistic
Riemann solvers recently developed for describing the evolution of special
relativistic flows. Our proposal relies on a local change of coordinates
in terms of which the spacetime metric is locally Minkowskian and permits
an accurate description of numerical general relativistic hydrodynamics.

\end{abstract}

\section{Introduction } \label{s:intro}

  Astrophysical scenarios involving relativistic flows have drawn the attention
and efforts of many researchers since the pioneering studies of May and White
\cite{MW67} and Wilson \cite{Wi72}. Relativistic jets, accretion
onto compact objects (in X-ray binaries or in the inner regions of active
galactic nuclei),
stellar core collapse, coalescing compact binaries (neutron star and/or black
holes) and recent models of formation of Gamma-ray bursts (GRBs) are examples of
systems in which the evolution of matter is described
within the frame of the theory of relativity (special or general).

  Since 1991 \cite{MIM91} the use of Riemann solvers in relativistic
hydrodynamics has proved successful in handling complex flows, with high Lorentz
factors and strong shocks, superseding traditional methods which
failed to describe ultrarelativistic flows \cite{NW86}. Exploiting the 
hyperbolic and conservative character of the relativistic hydrodynamical 
equations, we proposed how to extend {\it modern high-resolution 
shock-capturing} (HRSC) methods to the relativistic case, first 
in one-dimensional calculations  \cite{MIM91}, and, later on, we extended 
them to multidimensional special relativistic \cite{FIMM94} and
multidimensional general relativistic hydrodynamics \cite{Betal97}.
We made use of a linearized Riemann solver based on the {\it spectral 
decomposition} of the Jacobian matrices of the system. 

Unlike the case of classical fluid dynamics the use of HRSC techniques in 
the frame of relativistic fluid dynamics is very recent and has yet to cover 
the full set of possible applications. Up to now, the most interesting 
astrophysical applications have involved the simulation of extragalactic 
relativistic jets (see \cite{A99} and \cite{Ko99} for, respectively, 
relativistic 3D-hydro and 2D-magnetohydro calculations). 
Recently, some studies on the morphology of accreting flows onto moving black 
holes have being carried out (see \cite{fip99} and references therein)  
using a multidimensional general relativistic hydrocode. A very promising 
application of HRSC techniques in the frame of general relativistic 
magnetohydrodynamics has been used recently to simulate the formation of 
relativistic jets from black holes magnetized accretion disks \cite{KSK99}. 

  At present, to develop robust and accurate general relativistic hydrocodes is
a challenge in the field of Relativistic Astrophysics. A general relativistic
hydrocode is a useful research tool for studying flows which evolve in a
background spacetime. Furthermore, when appropriately coupled with Einstein
equations, such a general relativistic hydrocode is
crucial to model the evolution of matter in a dynamical spacetime. The coupling
between geometry and matter arises through the sources of the corresponding
system of equations. Such a marriage between numerical relativity and
numerical relativistic hydrodynamics could be useful, for
example, to analyze the dynamics (and the physics) of coalescing compact
binaries. These are one of the most promising sources of gravitational radiation
to be detected by the near future Earth-based laser-interferometer observatories 
of gravitational waves.

\section{The equations of general relativistic hydrodynamics
as a hyperbolic system of conservation laws}
\label{s:math}

The evolution of a relativistic fluid is governed by a system of equations 
which summarize {\it local conservation laws}: the local conservation of 
baryon number, $\nabla \cdot {\bf J} = 0 $, and the local conservation of 
energy-momentum, $\nabla \cdot {\bf T} = 0$ ($\nabla \cdot$ stands for the 
covariant divergence).


  If $\{ {\bf \partial_t, \partial_i} \}$ define the coordinate basis of
4-vectors which are tangents to the corresponding coordinate curves, then,
the {\it current of rest-mass}, $\bf J$,
and the {\it energy-momentum tensor}, $\bf T$, for a perfect fluid, have the
components:
$J^{\mu} = \rho u^{\mu}$, and
$T^{\mu\nu} = \rho h u^{\mu}u^{\nu} + p g^{\mu \nu}$, respectively,
$\rho$ being the rest-mass density, $p$ the
pressure and $h$ the specific enthalpy, defined by
$h = 1 + \varepsilon + p/\rho$, where
$\varepsilon$ is the specific internal energy. $u^{\mu}$ is the
four-velocity of the fluid and
 $g_{\mu \nu}$ defines the metric of
the spacetime $\cal M$ where the fluid evolves.
As usually, Greek (Latin) indices run from 0 to 3 (1 to 3)
-- or, alternatively, they stand for the general coordinates $\{t,x,y,z\}$
($\{x,y,z\}$) -- and the system of units is the
so-called geometrized $(c=G=1)$.

An equation of state $p=p(\rho,\varepsilon)$
closes, as usual, the system. Accordingly, the 
local sound velocity $c_{s}$ satisfies: $h c_{s}^{2} = \chi + (p/\rho^{2})\kappa$,
with $\chi = \partial p / \partial \rho |_{\epsilon}$ and
$\kappa = \partial p / \partial \epsilon|_{\rho}$.

Following \cite{Betal97}, let $\cal M$ be a general spacetime, described by the 
four dimensional metric tensor $g_{\mu \nu}$. According to the $\{3+1\}$ 
formalism, the metric is split into the objects $\alpha$ ({\it lapse}), 
$\beta^{i}$ ({\it shift}) and $\gamma_{ij}$, keeping the line element in the form:
\begin{equation}
ds^{2} = -(\alpha^{2}-\beta_{i}\beta^{i}) dt^{2}+
2 \beta_{i} dx^{i} dt + \gamma_{ij} dx^{i}dx^{j}
\end{equation}

If $\bf n$ is a unit timelike vector field normal to the
spacelike hypersurfaces $\Sigma_t$ (t = const.), then, by definition of
$\alpha$ and $\beta^i$ is:
${\bf \partial}_t = \alpha {\bf n} + \beta^i {\bf \partial}_i$,
with ${\bf n} \cdot {\bf \partial}_i$ = 0, $\,\, \forall i$.
Observers, ${\cal O}_{E}$, at rest in the slice $\Sigma_t$, i.e.,
those having ${\bf n}$ as four-velocity ({\it Eulerian observers}),
measure the following velocity of the fluid
\begin{equation}
v^i= \frac{u^i}{\alpha u^t} + \frac{\beta^i}{\alpha}
\end{equation}
\noindent
where $W \equiv -({\bf u \cdot n})= \alpha u^{t} $, the Lorentz factor, 
satisfies $W=(1-{\rm v}^{2})^{-1/2}$ with ${\rm v}^{2}= v_i v^i$ ($v_{i} = 
\gamma_{ij} v^j$). 

Let us define a basis adapted to the observer ${\cal O}_{E}$,
${\bf e}_{(\mu)}  = \{{\bf n}, {\partial}_i \}$,
and the following five four-vector fields 
$\{ {\bf J}$, ${\bf T}\cdot {\bf n}$, ${\bf T}\cdot \partial_1$, 
${\bf T}\cdot \partial_2$, ${\bf T}\cdot \partial_3 \}$.
Hence, the above system of equations of general relativistic hydrodynamics
(GRH) can be written
\begin{equation}
\nabla \cdot {\bf A} = s,
\label{diva}
\end{equation}
\noindent
where ${\bf A}$ denotes any of the above 5 vector fields, and $s$ is the 
corresponding source term.

The set of {\it conserved variables} gathers 
those quantities which are directly measured
by ${\cal O}_{E}$, i.e., the rest-mass density ($D$), the
momentum density in the $j$-direction ($S_j$) and the total energy density
($E$). In terms of the {\it primitive variables}
${\bf w} = (\rho, v_{i}, \varepsilon)$ ($v_{i} = \gamma_{ij} v^j$) they are
\begin{eqnarray}
D =  \rho W \,\,\,\,,\,\,\,\,
S_j  =  \rho h W^2 v_j \,\,\,\,,\,\,\,\,
E  =  \rho h W^2 - p
\end{eqnarray}

Taking all the above relations together, the fundamental
system to be considered for numerical applications is
\begin{equation}
\frac{1}{\sqrt{-g}} \left(
\frac {\partial \sqrt{\gamma}{\bf F}^{0}({\bf w})}
{\partial x^{0}} +
\frac {\partial \sqrt{-g}{\bf F}^{i}({\bf w})}
{\partial x^{i}} \right)
 = {\bf s}({\bf w})
\label{F}
\end{equation}
\noindent
where the quantities
${\bf F}^{\alpha}({\bf w})$ are
\begin{equation}
{\bf F}^{0}({\bf w})  =   (D, S_j, \tau)
\end{equation}
\begin{equation}
{\bf F}^{i}({\bf w})  =   \left(D \left(v^{i}-\frac{\beta^i}{\alpha}\right),
 S_j \left(v^{i}-\frac{\beta^i}{\alpha}\right) + p \delta^i_j,
\tau \left(v^{i}-\frac{\beta^i}{\alpha}\right)+ p v^{i} \right)
\end{equation}
\noindent
and the corresponding sources ${\bf s}({\bf w})$ are
\begin{eqnarray}
{\bf s}({\bf w}) =  \left(0,
T^{\mu \nu} \left(
\frac {\partial g_{\nu j}}{\partial x^{\mu}} -
\Gamma^{\delta}_{\nu \mu} g_{\delta j} \right),
\alpha  \left(T^{\mu 0} \frac {\partial {\rm ln} \alpha}{\partial x^{\mu}} -
T^{\mu \nu} \Gamma^0_{\nu \mu} \right)
                     \right)
\end{eqnarray}

\noindent
$\tau$ being
$\tau \equiv  E - D$, and 
$g\equiv \det(g_{\mu\nu})$ is such that
$\sqrt{-g} = \alpha\sqrt{\gamma}$ ($\gamma\equiv \det(\gamma_{ij})$)

\subsection{Linearized Riemann Solvers and Characteristic fields}

  Modern HRSC schemes use the characteristic structure of the hyperbolic system
of conservation laws. In many Godunov-type schemes, the characteristic structure
is used to compute either an exact or an approximate solution to a sequence of
Riemann problems at each cell interface. In characteristic based methods
the characteristic structure is used to compute the local
characteristic fields, which define the directions along which the
characteristic variables propagate.
In both these approaches, the characteristic decomposition of the Jacobian
matrices of the nonlinear system of conservation laws is important,
not only because it is one of the key ingredients in the
design of the numerical flux at the interfaces, but because
experience has shown that it facilitates a robust upgrading of
the order of a numerical scheme.

The three $5\times 5$-Jacobian matrices ${\bf \cal B}^{i}$
associated to system (\ref{F}) are
\begin{equation}
{\bf \cal B}^{i} = \alpha \frac{\partial{\bf F}^{i}}
{\partial {\bf F}^0}.
\label{B}
\end{equation}

The {\it eigenvalues} of ${\bf \cal B}^{x}$ are 
\begin{equation}
\lambda_0 = \alpha v^x - \beta^x \mbox{\,\,\,\,(triple)}
\label{lambda0}
\end{equation}
\noindent
which defines the {\it material waves}, and two other, $\lambda_{\pm}$,
associated with the {\it acoustic waves}
\begin{eqnarray}
\lambda_{\pm} = \frac{\alpha}{1-{v}^{2}c_{s}^{2}}
\left\{ v^x(1-c_{s}^{2}) \pm \right. \nonumber \\
              \mbox{} \left. {\pm} c_{s}
\sqrt{(1-{v}^{2}) [\gamma^{xx}(1-{v}^{2}c_{s}^{2})
-v^x v^x (1-c_{s}^{2})]} \right\} -  \beta^x
\label{lambdapm}
\end{eqnarray}

A complete set of {\it right-eigenvectors} is

\[
{\bf r}_{\pm} =
\left[ \begin{array}{c}
1 \\ \\
h W
\left( {\displaystyle{v_x - \frac{v^x - \Lambda_{\pm}^x}
{\gamma^{xx} - v^x \Lambda_{\pm}^x}
}}\right)  \\ \\
h W v_y   \\ \\
h W v_z   \\ \\
{\displaystyle{\frac{h W (\gamma^{xx} - v^x v^x)}
{\gamma^{xx} - v^x \Lambda_{\pm}^x}
}} -1  \\ \\
\end{array} \right] \,\,\,,\,\,\,
{\bf r_{0,1}} =
\left[ \begin{array}{c}
{\displaystyle{\frac{{\cal K}}{h W}}} \\ \\
v_x \\ \\
v_y \\ \\
v_z \\ \\
1 - {\displaystyle{\frac{{\cal K}}{h W}}} \\ \\
\end{array} \right]
\]

\[
{\bf r_{0,2}} =
\left[ \begin{array}{c}
W v_y \\ \\
h \left( {\displaystyle{\gamma_{xy} + 2 W^2 v_x v_y }}\right)  \\ \\
h \left( {\displaystyle{\gamma_{yy} + 2 W^2 v_y v_y }}\right)  \\ \\
h \left( {\displaystyle{\gamma_{zy} + 2 W^2 v_z v_y }}\right)  \\ \\
W v_y (2hW - 1)  \\ \\
\end{array} \right] \,\,\,,\,\,\,
{\bf r_{0,3}} =
\left[ \begin{array}{c}
W v_z \\ \\
h \left( {\displaystyle{\gamma_{xz} + 2 W^2 v_x v_z }}\right)  \\ \\
h \left( {\displaystyle{\gamma_{yz} + 2 W^2 v_y v_z }}\right)  \\ \\
h \left( {\displaystyle{\gamma_{zz} + 2 W^2 v_z v_z }}\right)  \\ \\
W v_z (2hW - 1)  \\ \\
\end{array} \right]
\]

The corresponding set of {\it left-eigenvectors} is

\begin{eqnarray*}
{\bf l}_{0,1} = {\displaystyle{\frac{W}{{\cal K} - 1}}} 
\left( h - W, W v^x, W v^y, W v^z, - W \right)
\end{eqnarray*}


\[
{\bf l}_{0,2} = {\displaystyle{\frac{1}{h \xi}}}
\left[ \begin{array}{c}
- \gamma_{zz} v_y + \gamma_{yz} v_z
\\  \\
v^x (\gamma_{zz} v_y - \gamma_{yz} v_z)
\\  \\
\gamma_{zz} (1 - v_x v^x) + \gamma_{xz} v_z v^x
\\  \\
- \gamma_{yz} (1 - v_x v^x) - \gamma_{xz} v_y v^x
\\  \\
- \gamma_{zz} v_y + \gamma_{yz} v_z
\end{array} \right] \,,\,
%
{\bf l}_{0,3} = {\displaystyle{\frac{1}{h \xi}}}
\left[ \begin{array}{c}
- \gamma_{yy} v_z + \gamma_{zy} v_y
\\  \\
v^x (\gamma_{yy} v_z - \gamma_{zy} v_y)
\\  \\
- \gamma_{zy} (1 - v_x v^x) - \gamma_{xy} v_z v^x
\\  \\
\gamma_{yy} (1 - v_x v^x) + \gamma_{xy} v_y v^x
\\  \\
- \gamma_{yy} v_z + \gamma_{zy} v_y
\end{array} \right]
\]
\[
{\bf l}_{\mp} = ({\pm} 1){\displaystyle{\frac{h^2}{\Delta}}}
\left[ \begin{array}{c}
h W {\cal V}^x_{\pm} \xi + l_{\mp}^{(5)}
\\  \\
\Gamma_{xx} (1 - {\cal K} {\tilde {\cal A}}^x_{\pm}) +
(2 {\cal K} - 1){\cal V}^x_{\pm} (W^2 v^x \xi - \Gamma_{xx} v^x)
\\  \\
\Gamma_{xy} (1 - {\cal K} {\tilde {\cal A}}^x_{\pm}) +
(2 {\cal K} - 1) {\cal V}^x_{\pm} (W^2 v^y \xi - \Gamma_{xy} v^x)
\\  \\
\Gamma_{xz} (1 - {\cal K} {\tilde {\cal A}}^x_{\pm}) +
(2 {\cal K} - 1) {\cal V}^x_{\pm} (W^2 v^z \xi - \Gamma_{xz} v^x)
\\  \\
(1 - {\cal K})[- \gamma v^x + {\cal V}^x_{\pm} (W^2 \xi - \Gamma_{xx})]
- {\cal K} W^2 {\cal V}^x_{\pm} \xi
\end{array} \right]
\]
\noindent
where the following auxiliary quantities and relations have been introduced:

\begin{eqnarray}
\Lambda_{\pm}^i \equiv {\tilde{\lambda}}_{\pm} + \tilde{\beta}^i
\,\,\,\,\,,\,\,\,\,\,
\tilde{\lambda} \equiv \lambda/\alpha
\,\,\,\,\,,\,\,\,\,\,
\tilde{\beta}^i \equiv \beta^i/\alpha
\end{eqnarray}
\begin{eqnarray}
{\cal K} \equiv {\displaystyle{\frac{\tilde{\kappa}}
{\tilde{\kappa}-c_s^2}}} \,\,\,\,\,,\,\,\,\,\,
\tilde{\kappa}\equiv \kappa/\rho
\end{eqnarray}
\begin{eqnarray}
{\cal C}^x_{\pm} \equiv v_x - {\cal V}^x_{\pm} \,\,\,,\,\,\,
{\cal V}^x_{\pm} \equiv {\displaystyle{\frac{v^x - \Lambda_{\pm}^x}
{\gamma^{xx} - v^x \Lambda_{\pm}^x}}}
\end{eqnarray}
\begin{eqnarray}
{\tilde {\cal A}}^x_{\pm} \equiv
{\displaystyle{\frac{\gamma^{xx} - v^x v^x}
{\gamma^{xx} - v^x {\Lambda}^x_{\pm}}}}
\end{eqnarray}
\begin{eqnarray}
\Delta \equiv h^3 W ({\cal K} - 1) ({\cal C}^x_{+} - {\cal C}^x_{-} ) \xi
\end{eqnarray}
\begin{eqnarray}
\xi \equiv \Gamma_{xx}  - \gamma v^x v^x
\,\,\,,\,\,\, \Gamma_{xx} = \gamma_{yy} \gamma_{zz} - \gamma_{yz}^2
\,\,\,,\,\,\, ...
\end{eqnarray}

Symmetry arguments allow one to obtain the spectral decomposition in the other 
spatial directions. The corresponding expressions in Special Relativity 
\cite{dfim98} are easily covered. The above full spectral decomposition provides 
the user with the technical ingredients needed to develop state-of-the-art, 
upwind-based HRSC codes for numerical relativistic hydrodynamics.
In 3D general-relativistic applications, and depending on the particular
Riemann solver or flux formula used, the knowledge of the analytical
values of the left-eigenvectors could be crucial in 
the efficiency of the code (see \cite{api99}).


\section{Special Relativistic Riemann Solvers in 
General Relativistic Hydrodynamics}
\label{s:srrs}

Up to now, only a small number of papers have considered the
extension of HRSC methods to GRH using linearized Riemann Solvers
or flux formulae (\cite{MIM91} and \cite{Retal96} for 1D problems, 
\cite{Betal97}, \cite{FMST99} and \cite{papadopoulos99}).
or deriving explicitly an
extension of Roe's Riemann solver to GRH
(\cite{EM95}).

In \cite{PFIMM98} we have answered the following basic question
(suggested in \cite{Balsara94} and \cite{Marti97}): 
Is it possible to obtain a general procedure that allows one to take 
advantage of Special Relativistic Riemann Solvers (SRRS) to generate numerical 
solutions describing the evolution of relativistic flows in strong gravitational 
fields?

The affirmative answer relies on a {\it local change of coordinates}, at 
each numerical interface, in terms of which the spacetime metric is locally 
Minkowskian. Our procedure, hence, follows analogous trends to those used 
in classical fluid dynamics to solve Riemann problems in general curvilinear 
coordinates.
  
Let us consider the integral form of the system of equations (\ref{diva}) 
on a four--dimensional volume $\Omega$, with three--dimensional 
boundary ${\partial \Omega}$, and apply Gauss theorem to obtain the 
corresponding balance equation

\begin{equation}
\int_{\partial \Omega} {\bf A} \cdot  d^3{\bf \Sigma} 
= \int_{\Omega} s d\Omega.
\label{intdiva}
\end{equation}

For numerical applications, we choose volume $\Omega$ as the one bounded by
the coordinate hypersurfaces $\{\Sigma_{x^{\alpha}}$, $\Sigma_{x^{\alpha}+
\Delta x^{\alpha}}\}$. Hence, the time variation of the mean value of $A^0$,
\begin{equation}
\displaystyle{\overline A^0} = \frac{1}{\Delta {\cal V}}
\int^{x^1+\Delta x^1}_{x^1}
\int^{x^2+\Delta x^2}_{x^2} \int^{x^3+\Delta x^3}_{x^3}
\sqrt{-g} \, A^0 dx^1 dx^2 dx^3,
\end{equation}  
within the spatial volume 
\begin{equation}
{\Delta {\cal V}} = \int^{x^1+\Delta x^1}_{x^1} 
\int^{x^2+\Delta x^2}_{x^2} \int^{x^3+\Delta x^3}_{x^3} \sqrt{-g} 
dx^1 dx^2 dx^3 ,
\end{equation}
can be obtained from
\begin{eqnarray}
(\displaystyle{\overline A^0} {\Delta {\cal V}})_{t+\Delta t} = &
(\displaystyle{\overline A^0} {\Delta {\cal V}})_{t} + 
\displaystyle{\int_{\Omega}} s d\Omega \, - 
\nonumber \\ &
\left(
\displaystyle{\int_{\Sigma_{x^1}}} {\bf A} \cdot d^3{\bf \Sigma} +
\displaystyle{\int_{\Sigma_{x^1+\Delta x^1}}}{\bf A} \cdot d^3{\bf \Sigma} + 
\right.
\nonumber \\ &
\,\,\displaystyle{\int_{\Sigma_{x^2}}}{\bf A} \cdot d^3{\bf \Sigma} +
\displaystyle{\int_{\Sigma_{x^2+\Delta x^2}}}{\bf A} \cdot d^3{\bf \Sigma} + 
\nonumber \\ &
\left.
\,\displaystyle{\int_{\Sigma_{x^3}}}{\bf A} \cdot d^3{\bf \Sigma} +
\displaystyle{\int_{\Sigma_{x^3+\Delta x^3}}}{\bf A} \cdot d^3{\bf \Sigma} 
\right).
\label{integ}
\end{eqnarray}
\noindent

In order to advance in time, the volume and surface integrals on the
right hand side have to be evaluated. 
We have applied HRSC to calculate
the ${\bf A}$ vector fields by solving local 
Riemann problems combined with monotonized cell reconstruction 
techniques.


According to the Equivalence Principle, physical laws in a {\it local inertial 
frame} of a curved spacetime have the same form as in special relativity
(see, e.g., Schutz \cite{Sch85}). 
Locally flat (or geodesic) systems of coordinates, 
in which the metric is brought into the Minkowskian form up to second order 
terms, are the realization of these local inertial frames. However, 
whereas the coordinate transformation leading to locally flat
coordinates involves second order terms, 
locally Minkowskian coordinates 
are obtained by a linear transformation.
This fact is of crucial importance when exploiting the self-similar 
character of the solution of the Riemann problems 
set up across the coordinate surfaces.
 
Hence, we propose to perform
a coordinate transformation 
to locally Minkowskian coordinates at each numerical interface assuming 
that the solution of the Riemann problem is one in special 
relativity and planar symmetry. 
This last assumption 
is equivalent to the approach followed in classical fluid 
dynamics, when using the solution of Riemann problems in slab symmetry for 
problems in cylindrical or spherical coordinates, which breaks down
near the singular points ({\it e.g.} the polar axis in cylindrical
coordinates). 
Analogously to classical fluid dynamics, 
the numerical error will depend on the magnitude of 
the Christoffel symbols, which might be large whenever
huge gradients or large temporal variations 
of the gravitational field are present. Finer grids and improved
time advancing methods will be required in those regions.

In the rest of this section we will focus on the evaluation of the 
first flux integral in Eq. (\ref{integ}).
\begin{eqnarray}
\label{intg1}
\int_{\Sigma_{x^1}} {\bf A}\cdot d^3{\bf \Sigma}=
\int_{\Sigma_{x^1}}{A^{1}}\sqrt{-g}\,dx^0dx^2dx^3 
\end{eqnarray}
To begin, we will express the integral on a 
basis ${\bf e}_{\hat{\alpha}}$ with
${\bf e}_{\hat{0}} \equiv {\bf n}$ and
${\bf e}_{\hat{i}}$ forming an orthonormal 
basis in the plane orthogonal to ${\bf n}$ with ${\bf e}_{\hat{1}}$
normal to the surface $\Sigma_{x^1}$ and
${\bf e}_{\hat{2}}$ and ${\bf e}_{\hat{3}}$ tangent to that surface.
The vectors of this basis verify
${\bf e}_{\hat{\alpha}}\cdot {\bf e}_{\hat{\beta}} = 
\eta_{\hat{\alpha}\hat{\beta}}$
with $\eta_{\hat{\alpha}\hat{\beta}}$ the Minkowski metric (in the
following, caret subscripts will refer to vector 
components in this basis).

Denoting by $x^\alpha_0$ the coordinates at the center of the interface
at time $t$, we introduce the following locally Minkowskian coordinate system
\begin{equation}
x^{\hat{\alpha}} = \left.M\right._\alpha^{\hat{\alpha}}(x^\alpha-x^\alpha_0), 
\label{change}
\end{equation}
where the matrix $M_\alpha^{\hat{\alpha}}$ is given by
$\partial_{\alpha}=M_{\alpha}^{\hat{\alpha}} {\bf e}_{\hat{\alpha}}$, 
calculated at $x^\alpha_0$.
In this system of coordinates the equations of general relativistic 
hydrodynamics transform into the equations of special relativistic 
hydrodynamics, in Cartesian coordinates,
but with non-zero sources, and the flux integral (\ref{intg1}) reads
\begin{eqnarray}
\label{intg2}
\int_{\Sigma_{x^1}}(A^{\hat{1}} - \frac{\beta^{\hat{1}}}
{\alpha}A^{\hat{0}})
\sqrt{-\hat{g}}\,dx^{\hat{0}}dx^{\hat{2}}dx^{\hat{3}} 
\end{eqnarray}
with $\sqrt{-\hat{g}} = 1+ {\cal O} (x^{\hat{\alpha}})$,
where we have taken into account that, in the coordinates 
$x^{\hat{\alpha}}$, 
$\Sigma_{x^1}$ is described by the equation $x^{\hat{1}} - 
\frac{\beta^{\hat{1}}}
{\alpha} x^{\hat{0}} = 0$ (with $\beta^{\hat{i}} = M^{\hat{i}}_i \beta^i$),
where the metric elements ${\beta^{1}}$ and ${\alpha}$ are calculated 
at $x^{\alpha}_0$.
Therefore, this surface is not at rest but moves with {\it speed} 
$\,\,{\beta^{\hat{1}}}/{\alpha}$.

At this point, all the theoretical work on SRRS developed in recent years, can 
be exploited.  
The quantity in parenthesis in (\ref{intg2}) represents the numerical
flux across $\Sigma_{x^1}$, which can,
now, be calculated by solving the special relativistic Riemann problem
defined with the values at the two sides of $\Sigma_{x^1}$
of two independent thermodynamical 
variables (namely, the rest mass density $\rho$ and the specific internal 
energy $\epsilon$) and the components of the velocity in the orthonormal 
spatial basis $v^{\hat{i}}$ ($v^{\hat{i}}=M_i^{\hat{i}} v^i$).
Although most linearized Riemann solvers provide the numerical fluxes 
for surfaces at rest, it is easy to apply them to moving 
surfaces, relying on the conservative and hyperbolic character 
of the system of equations (as in \cite{HH83}).

Once the Riemann problem has been solved,
by means of any linearized or exact SRRS, we can take advantage
of the self-similar character of the solution of the
Riemann problem, which makes it constant
on the surface $\Sigma_{x^1}$ simplifying
the calculation of the above integral enormously (\ref{intg2}):

\begin{equation}
\label{intg3}
\int_{\Sigma_{x^1}} {\bf A}\cdot d^3{\bf \Sigma}=
(A^{\hat{1}} - \frac{\beta^{\hat{1}}}
{\alpha}A^{\hat{0}})^{*}\int_{\Sigma_{x^1}}
\sqrt{-\hat{g}}\,dx^{\hat{0}}dx^{\hat{2}}dx^{\hat{3}} 
\end{equation}
\noindent
where the superscript (*) stands for the value 
on $\Sigma_{x^1}$ obtained from the solution
of the Riemann problem.
The integral in the right hand side of (\ref{intg2}) is the 
area of the surface $\Sigma_{x^1}$ and can be expressed in terms
of the original coordinates as
\begin{equation}
\label{intg4}
\int_{\Sigma_{x^1}}\sqrt{\gamma^{11}}
\sqrt{-{g}}\,dx^{{0}}dx^{{2}}dx^{{3}} 
\end{equation}
which can be evaluated for a given metric.

  Finally, notice that the numerical fluxes defined in (\ref{intg2}) correspond 
to the vector fields ${\bf A} = \{ {\bf J}$, ${\bf T}\cdot {\bf n}$, ${\bf 
T}\cdot e_{\hat {1}}$, ${\bf T}\cdot e_{\hat {2}}$, ${\bf T}\ e_{\hat {3}} \}$.
Thus the additional relation 
\begin{equation}
{\bf T}\cdot \partial_i=M_{i}^{\hat{j}}({\bf T}\cdot {\bf e}_{\hat{j}}) 
\end{equation}
has to be used for the momentum equations.

The interested reader can address reference \cite{PFIMM98} for
details on the testing and calibration of our procedure.
The additional computational cost of the approach is completely
negligible. The procedure has a large potentiality and can be applied to
other systems of conservation laws, such as magneto-hydrodynamics (MHD), 
making possible to solve the general relativistic MHD equations
using the corresponding Riemann solvers developed for the special 
relativistic case.

\section{Conclusions} \label{s:concl}


An appropriate conservative formulation for the equations, together with
the knowledge of the characteristic fields associated to the system, define 
the starting point for using HRSC schemes. The spectral decomposition of the 
Jacobian matrices, corresponding to the fluxes in each spatial direction, is 
used in the numerical flux computation and, moreover, it is potentially
interesting in allowing an extensive range of application of HRSC methods 
with different approximate Riemann solvers or flux formulae. 

The procedure outlined in  Section $\S 3$
is --from the computational point of view -- very cheap, since
it involves a linear change of coordinates. It
has a large potentiality and can be 
applied to other systems of conservation laws, as magneto-hydrodynamics,
giving a very useful numerical tool to solve the general
relativistic MHD equations using the corresponding Riemann solvers
developed for the special relativistic case.
In particular, it is possible to use the exact solution 
of the special relativistic Riemann problem \cite{MM94}, \cite{Pons99}. 

The astrophysical applications foreseen in the present and near 
future include the study of jet formation, multidimensional stellar core 
collapse, gamma-ray bursts and the coalescence of compact binaries. HRSC 
methods can, without doubt, be used successfully to tackle these scenarios, 
and acquire the prestige they have already earned in the 
simulation of relativistic jets and accretion flows around compact objects.


\paragraph{\it Acknowledgments:} This work has been supported by the Spanish
DGES (grant PB97-1432).  J.A.F. acknowledges financial support from a TMR 
fellowship of the European Union (contract nr. ERBFMBICT971902).

%


\begin{thebibliography}{99}
%

\bibitem{A99}
Aloy M.A. (1999),
this volume.

\bibitem{api99}
Aloy M.A., Pons J.A., and
Ib{\'a}{\~n}ez J.M$^{\underline{\mbox{a}}}$. (1999),
{\it Comp. Phys. Comm.}, {\bf 120}: 115.

\bibitem{Balsara94} 
Balsara D.S. (1994),
{\it J. Comput. Phys.}, {\bf 114}: 284.

\bibitem{Betal97}
Banyuls F., Font J.A.,
Ib{\'a}{\~n}ez J.M$^{\underline{\mbox{a}}}$.,
Mart\'{\i} J.M$^{\underline{\mbox{a}}}$., and Miralles J.A. (1997),
{\it ApJ}, {\bf 476}: 221.


\bibitem{dfim98}
Donat R., Font J.A., 
Ib{\'a}{\~n}ez J.M$^{\underline{\mbox{a}}}$.,
and Marquina A. (1998),
{\it J. Comput. Phys.}, {\bf 146}: 58.


\bibitem{EM95}
Eulderink F., and Mellema G. (1995),
{\it A\&A Suppl.}, {\bf 110}: 587.

\bibitem{FIMM94}
Font J.A., Ib\'a\~{n}ez J.M$^{\underline{\mbox{a}}}$., 
Marquina A., and 
Mart\'{\i} J.M$^{\underline{\mbox{a}}}$. (1994), 
{\it A\&A}, {\bf 282}: 304.

\bibitem{fip99}
Font J.A., Ib{\'a}{\~n}ez J.M$^{\underline{\mbox{a}}}$., and
Papadopoulos P. (1999),
this volume.

\bibitem{FMST99}
Font J.A., Miller M., Suen W.-M., and Tobias M. (1999),
{\it Phys. Rev. D}, in press (gr-qc/9811015).

\bibitem{HH83} 
Harten A., and Hyman J.M. (1983), 
{\it J. Comput. Phys.}, {\bf 50}: 235.

\bibitem{Ko99}
Komissarov S. (1999),
this volume.

\bibitem{KSK99}
Koide S., Shibata K., and Kudoh T. (1999), 
{\it ApJ}, {\bf 522}: 727.

\bibitem{Marti97}
Mart\'{\i} J.M$^{\underline{\mbox{a}}}$. (1997),
in {\it Relativistic Gravitation and Gravitational
Radiation}, Proceedings of the Les Houches School of
Physics (26 September-6 October, 1995), ed. by
Marck J-A., and Lasota J-P., Cambridge University Press.

\bibitem {MIM91}
Mart\'{\i} J.M$^{\underline{\mbox{a}}}$.,
Ib\'a\~{n}ez J.M$^{\underline{\mbox{a}}}$., and Miralles J.A. (1991), 
{\it Phys. Rev. D}, {\bf 43}: 3794.

\bibitem {MM94}
Mart\'{\i} J.M$^{\underline{\mbox{a}}}$., and M\"uller E. (1994),
{\it J. Fluid Mech.}, {\bf 258}: 317.

\bibitem{MW67}
May M.A., and White R.H. (1967), 
{\it Math. Comp. Phys.}, {\bf 7}: 219.

\bibitem{NW86}
Norman M.L., and Winkler K-H.A. (1986), 
in {\it Astrophysical Radiation Hydrodynamics}, ed. by
Norman M.L., and Winkler K-H.A., Reidel Publ.

\bibitem{papadopoulos99}
Papadopoulos P., and Font J.A. (1999),
{\it Phys. Rev. D}, in press \\
\noindent
(gr-qc/9902018).

\bibitem{Pons99}
Pons J.A., 
Mart\'{\i} J.M$^{\underline{\mbox{a}}}$., and M\"uller E. (1999),
this volume.

\bibitem{PFIMM98}
Pons J.A., Font J.A.,
Ib\'a\~{n}ez J.M$^{\underline{\mbox{a}}}$., 
Mart\'{\i} J.M$^{\underline{\mbox{a}}}$.,
and Miralles J.A. (1998), 
{\it A\&A}, {\bf 339}: 638.

\bibitem{Retal96}
Romero J.V., Ib\'a\~{n}ez J.M$^{\underline{\mbox{a}}}$.,
Mart\'{\i} J.M$^{\underline{\mbox{a}}}$., and Miralles J.A. (1996),
{\it ApJ}, {\bf 462}: 839.

\bibitem{Sch85} 
Schutz B.F. (1985),
{\it A first course in general relativity}, 
Cambridge University Press.




\bibitem {Wi72}
Wilson J.R. (1972), 
{\it ApJ}, {\bf 173}: 431.


\end{thebibliography}
\end{document}